\documentclass[11pt]{amsart}

\usepackage{amssymb,amsmath,amscd}
\usepackage{bbm}
\usepackage{a4wide}

\theoremstyle{plain}

\theoremstyle{definition}

\newcommand{\ts}{\hspace{0.5pt}}

\begin{document}

\title[Aperiodicity in equilibrium systems:
 Between order and disorder. ]
{Aperiodicity in equilibrium systems: \\
Between order and disorder \\[2mm]
}


\author{Aernout C.D. van Enter}
\address{
Johann Bernoulli Institute for Mathematics and
  Computer Science, \newline
 \hspace*{\parindent}University of Groningen, 
  PO Box 407, 9700\ts AK Groningen, 
The Netherlands}
\email{a.c.d.van.enter@rug.nl}

\maketitle

{\bf Abstract:}

Spatial aperiodicity occurs in various models and materials. Although today the most well-known examples occur in the area of quasicrystals, other applications might also be of interest. Here we discuss some issues related to the notion and occurrence of aperiodic order in equilibrium statistical mechanics. In particular, we consider  some spectral characterisations, and shortly review what is known about the occurrence of aperiodic order in lattice models at zero and nonzero temperatures. At the end some more speculative connections to the theory of (spin-)glasses are indicated.

\smallskip
{\bf PACS numbers:} 05.70.Fh, 61.44.Br, 02.50.Ey

\bigskip 
\section{Some Questions}
Aperiodicity (much of the mathematics of which is discussed in the recent 
\cite{BG}) describes  the quasiperiodic order of quasicrystals, as well as 
weaker forms of long-range order. In fact a very famous   early, rather 
speculative, mentioning, of ``aperiodic crystals'' already occurred in 
Schr\"odinger's ''What is Life?'' \cite{Sch}. That aperiodicity was a presumed property of chromosomes. The notion of aperiodicity gives rise to various
questions, three of which I will discuss here.  

\smallskip

{\bf Question 1}: Aperiodic order as a form of order. How should one describe ``general'' long-range order in equilibrium systems? In particular how can one give  spectral characterisations thereof, which will be visible in some kind  of diffraction spectrum?  

\smallskip

{\bf Question 2}: Where does aperiodic order come from? Can one find statistical mechanical models of quasicrystals or ``weak crystals'' in which either ground states or Gibbs states display aperiodic order? Is it necessary to use long-range interactions or can one also obtain this behaviour with finite-range interactions? How does this depend on the dimension?  

\smallskip

{\bf Question 3}: Aperiodic order as a form of disorder. 
What can findings in the quasicrystal community teach us about theory of (spin-) glasses \cite{Kur}?\\
E.g.:
How (dis)ordered can  tilings and tiling models be? Can we learn from them about materials other than quasicrystals, such as e.g. (spin) glasses?
Or can one compute and obtain  aperiodic examples which display an interesting  
Parisi overlap distribution (a quantity which was introduced for the 
paradigmatic disordered model, spin glasses \cite{NSbook})?  

\smallskip

In this contribution I will discuss the above  questions in the technically simple context of statistical mechanical lattice models. 
Although a lot of the theory of aperiodic order has been developed looking at individual samples, it makes sense to consider probability measures (ensembles) on those. Ergodic arguments tell us that the behaviour of all typical -probability one- samples will be the same, once the probability measures are spatially ergodic, and then results easily carry over. Such properties are said to be ``self-averaging''. (In the case when the probability measure is strictly ergodic, that is, it is uniquely ergodic --there exists only one translation invariant measure--and also  minimal --every orbit is dense--, it is even true that {\it all} samples behave the same. In other words, whereas the ergodic theorem in general implies that all translation-invariant sets have measure zero or one, in the strictly ergodic case, such sets are either the whole space or the empty set.) For some early papers developing this point of view for diffraction questions, see \cite{Dw,EM}.\\    

\smallskip

We consider configuration spaces 
$\Omega = {\Omega_0}^{Z^d}$, where the single-site space $\Omega_0$ is finite, whether 2-valued or 
many-valued, as happens in tiling models. Configurations, which are the elements of the configuration space,  are denoted by $\omega$'s or $\sigma$'s. On those $\Omega$'s we consider 
translation-invariant probability measures which are either ground states -for 
temperature zero- or Gibbs measures -at positive temperatures- for some interaction. A configuration often is viewed as a translation bounded measure (e.g. a Dirac comb) \cite{Hof}.\\
Interactions will be translation invariant and are thus given by a translation 
covariant collection of functions $\Phi_X(\omega), X \subset Z^d$, on 
${\Omega_0}^{X}$. That is, $\Phi_{\tau_i X}(\omega)= \Phi_X(\tau_{-i} \omega)$, where $\tau_i$ denotes a translation over an arbitrary  lattice vector $i$. The set of $\Phi$ will satisfy some summability condition of the form 

\begin{equation}
\sum_{0 \in X} ||\Phi_{X}|| g(X) \lneq \infty
\end{equation}
for some translation-invariant real function $g(X)$ defined on the finite subsets of $Z^d$, where $||.||$ denotes a supremum norm.\\
If the $g(X)$ in the summability condition grows fast enough with either the diameter or the number of sites in $X$, this implies a decay property of the interaction $\Phi$. \\
Gibbs measures for interactions are probability measures on $\Omega$.
Their conditional probabilities of finding configurations $\sigma_{\Lambda}$ in a finite $\Lambda$, given boundary condition (external configuration) $\omega^{\Lambda^{c}}$  are of Gibbsian form for the local energy (Hamilton) function  
$H^{\omega}_{\Lambda}(\sigma^{\Lambda})$.  
Here $H^{\omega}_{\Lambda}(\sigma^{\Lambda})= \sum_{X \cap \Lambda\neq \emptyset} \Phi_{X}(\sigma^{\Lambda} \omega^{\Lambda^c})$. 
Thus 
\begin{equation}
\mu^{\alpha}_{\Lambda}(\sigma^{\Lambda}|\omega^{\Lambda^c})= \frac{exp ( -\beta  H^{\omega}_{\Lambda}(\sigma^{\Lambda}))}{Z_{\Lambda}^{\omega}}, 
\end{equation}
for each choice of $\Lambda, \sigma$, and $\omega$, and each Gibbs measure $\mu^{\alpha}$. Here the inverse temperature is given by $\beta$.\\
 As long as $g$ in the summability condition is larger than a constant, one can define ground state measures and Gibbs measures for $\Phi$, see e.g. \cite{Geo,EFS}. It is always possible to find translation-invariant examples of those measures, for translation-invariant interactions. 

\smallskip

A traditional definition of ``order'' rests on the existence  of more than one  Gibbs measure for a given interaction. In such cases there exist  correlation functions which do not decay. In the case of the Ising ferromagnet, for example (equivalent to an attractive lattice gas), there exist two extremal translation-invariant Gibbs measures, the ``plus'' and the ``minus'' one, at sufficiently low temperatures. If one considers $\mu$ to be the Gibbs measure which is a symmetric convex combination (the average) of these two, the pair correlations do not asymptotically factorise, that is they do not converge to zero. In the case of aperiodic order, often there exist many Gibbs measures, but only one translation-invariant one, which is a mixture (convex combination) of the non-translation-invariant ones, with asymptotically non-factorising correlations. This then will be the Gibbs measure to consider.  

\smallskip

The quantities to consider for diffraction questions are the pair correlation functions 
\begin{equation}
f(n)=\mu (\sigma^0 \sigma^n)
\end{equation}
and their Fourier transforms, which in general are measures on $R^d$ (or on $d-$dimensional tori).

For the Parisi distribution, one needs to consider 
the overlap between two configurations $\sigma_1$ and $\sigma_2$, which 
is given by
\begin{equation}
q(\sigma_1, \sigma_2) = lim_{\Lambda \to \infty} \frac{1}{|\Lambda|} \sum_{i\in  \Lambda} {\sigma_1}^i {\sigma_2}^i.
\end{equation}
Its distribution then is computed with respect to the product measure of the 
system under consideration, that is the product of a ground state or Gibbs 
measure with itself.

\section{Some Partial Answers}

As for Question 1, in \cite{EM} we  investigated the distinction between what is now called ``Diffraction versus Dynamical Spectrum'', 
with an interpretation in terms of  atomic versus molecular long-range order. 
It  became  well-known afterwards, based on work by Baake, Lee, Lenz, Moody, Schlottmann and Solomyak,
\cite{BLM,LMS,Martin,BL,LS} 
that pure point 
diffraction and pure point dynamical spectrum, under some mild assumptions, are
equivalent properties of dynamical systems of translation bounded
measures.  But this type of equivalence does not extend to
systems with continuous spectrum, as the example of the Thue-Morse
sequences first showed \cite{EM}. 
The diffraction spectrum  of the Thue-Morse system is  singular 
continuous, while the dynamical spectrum has a non-trivial pure point part in 
form of the dyadic rationals. This spectral
information is not reflected in the diffraction spectrum.
However, this `missing' part can be extracted from the diffraction of a factor, 
 the so-called period doubling sequences, which are
Toeplitz sequences. In \cite{BE} an even simpler 
example of this phenomenon was presented for a one-dimensional system of 
random dimers, which can be of $+-$ 
or $-+$ type, and which can be located on $[2n, 2n+1]$ or  on $[2n,2n-1]$ intervals.
 This system has absolutely continuous diffraction spectrum, but the long-range 
order associated to the location of the dimers --''odd'' or ``even''-- provides an additional point in 
the dynamical spectrum. It appears thus that molecules can be more, but not less, ordered than their constituent atoms. For a more general analysis, indicating how the dynamical spectrum can be obtained as the union of the spectra of various (ergodic) factors of a system, see \cite{BEL}.\\

\smallskip

It is unknown whether the statement that molecules can be more but not less ordered than atoms extends to the case of singular spectrum; in particular, if the Diffraction Spectrum has no absolutely continuous component, does the same hold true for the Dynamical Spectrum?\\ 
It should be noted, that, despite the official characterisation of ``crystals'' in terms of their discrete spectrum, inspired by the famous quasicrystal discovery of Schechtman \cite{BCGS}, both before and after his discovery, alternative and more general notions of long-range order in ``weak'' or ``turbulent'' crystals have been proposed \cite{Rue, Rue2, EM,BenA}. We also note that next to the discrete spectrum indicating long-range order, some absolutely continuous spectrum, due to the existence of thermal fluctuations is expected. For some examples where either thermal or independent fluctuations contribute a continuous spectrum component, see \cite{BBM,BH,BS,BZ,Kue1,Kue2}. 

\smallskip

As for Question 2,  one can construct aperiodic tilings which are 
ground states for nearest-neighbor 
(tiling) models, by associating a positive energy to nearest-neighbor pair of tiles which violates the matching rules, and zero  energy when the matching rules are satisfied.
In one dimension one can choose aperiodic sequences, which can be shown to be 
ground states for long-range (lattice) interactions. Some stability and 
intrinsic frustration properties for tiling models have been proven, but as for 
positive temperatures (Gibbs states) one is till now restricted to 
one-dimensional aperiodic long-range order, which occurs for infinite-range 
interactions. 
This can occur either for one-dimensional long-range models, or for 
exponentially decaying interactions which are stabilised in two other 
directions, see e.g. \cite{Aub,EM2,EMZ,EZ,GERM,Mi1,Mi2,Mi3,Mi4, Rad,Rad2}. For 
many-body interactions the existence of ``rigid'' or ``frozen''  
aperiodic long-range order was recently proven in \cite{BLep}.\\   
To prove the existence of a properly quasicrystalline state for a finite-range model remains an open question, however. There exists a conjecture,
that for finite-range models in one and two dimensions at finite temperature there can occur only finitely many extremal Gibbs measures, see e.g. \cite{Slaw}. This would imply that  quasicrystalline order cannot appear below three dimensions. The conjecture, however, is not believed by everyone \cite{ArRad}.

\smallskip 

As for Question 3, tilings and sequences are known to exist with only absolutely continuous (diffraction or dynamical) spectrum, -even uncorrelated sequences and tilings-, which have zero entropy, as for example occurs in the Rudin-Shapiro system. Having zero entropy is a very weak form of ``order'', which can go together with having no order at all in the spectral sense, see e.g. \cite{BG2,KuLev} for a discussion of this point.

As regards overlap distributions, in \cite{EHM} it was observed that continuous (absolutely or singular continuous, or a combination of the two)  diffraction 
spectrum implies a trivial overlap distribution, whereas the Fibonacci sequences
provide an example with a continuous overlap distribution and the 
period-doubling Toeplitz sequences have a discrete, ultrametric overlap 
distribution. More recently, in \cite{EG} this was extended, 
to show that continuous overlap distributions occur for general Sturmian 
sequences (= ``balanced words`` = ''most homogeneous configurations''), 
and moreover, for paperfolding sequences a discrete ultrametric overlap 
distribution with dense support was found. For a related observation on Fibonacci and Sturmian systems, see also \cite{KuLev}.\\
Although in the theory of spin glasses a huge progress has occurred for 
mean-field models of the Sherrington-Kirkpatrick type (due especially to Guerra 
and Talagrand \cite{Pan,Tal,NSbook}), not much is known about short-range models. Thus aperiodic 
examples may play a useful role in illustrating various possibilities. 
E.g.,  the fact that the overlap distribution is disorder-independent becomes 
much more plausible once one realises that one does not need disorder at all 
to obtain nontrivial overlap distributions. 

It would be interesting to obtain examples also in higher dimensions, via tiling constructions. 

It should be mentioned here that tiling models for the glass transition have been investigated  in e.g. \cite{LePa,LevW,KuLev,NHL,Sas1,Sas2}. This seems to be a promising new direction to explore the role of aperiodic order in a new context.

Summarising,  spatial aperiodicity has a large role to play in condensed matter and mathematical physics, even beyond the description of quasicrystals.   

\bigskip

{\bf Acknowledgements:} I wish to thank the conference organisers of ICQ12 for inviting me, and  my coauthors for all they taught me.

\end{document}